# Experimental demonstration of dynamical input isolation in nonadiabatically modulated photonic cavities

Avik Dutt[1], Momchil Minkov[1], Qian Lin[2], Luqi Yuan[1], David A. B. Miller[1] & Shanhui Fan[1,*]

[1]Ginzton Laboratory and Department of Electrical Engineering, Stanford University, Stanford, CA 94350.

[2]Department of Applied Physics, Stanford University, 348 Via Pueblo, Stanford, CA 94305.

## ABSTRACT

Modulated optical cavities have been proposed and demonstrated for applications in communications, laser frequency stabilization, microwave-to-optical conversion and frequency comb generation. However, most studies are restricted to the adiabatic regime, where either the maximum excursion of the modulation or the modulation frequency itself is below the linewidth of the cavity. Here, using a fiber ring resonator with an embedded electro-optic phase modulator, we investigate the nonadiabatic regime. By strongly driving the modulator at frequencies that are significantly smaller than the free-spectral range of the ring resonator, but well beyond the linewidth of the resonator, we experimentally observe counterintuitive behavior predicted in a recent theoretical study by Minkov et al. [APL Photonics 2, 076101 (2017)], such as the complete suppression of drop-port transmission even when the input laser wavelength is on resonance with the optical cavity. This can be understood as dynamical isolation of the cavity from the input light. We also show qualitative differences in the steady-state responses of the system between the adiabatic and nonadiabatic limits. Our experiments probe a seldom explored regime of operation that is promising for applications in integrated photonic systems with current state-of-the-art technology.

Dynamically modulated cavities are important for applications such as optical communications,[1–4] quantum state transfer between microwave and optical photons,[5,6] and signal processing[3,7,8]. Specifically, dynamically modulated ring resonators, where the refractive index of a portion of the ring is varied periodically in time, have been predominantly explored in two qualitatively different regimes. In the first regime, the modulation is at or near the free spectral range (FSR) of the cavity to couple multiple longitudinal frequency modes. This regime finds use in active modelocking of ultrashort pulses,[9] generating frequency combs,[10–14] and in realizing synthetic frequency dimensions[15–20]. In the second regime, the modulation frequency is far below the FSR, and this finds applications in realizing efficient and compact electro-optic modulators (EOMs)[1,4,21–24]. Here, a single-mode description of the cavity suffices. Previous work in this case has almost exclusively focused on the adiabatic regime, which corresponds to the case where the modulation strength is sufficiently small, and the modulation frequency is less than the cavity linewidth. For a static resonator, the transport properties of light through a resonator are determined by the resonant frequency. In the adiabatic regime, the transport properties of the resonators are determined by the instantaneous resonant frequency.

Recent theoretical work predicts that for the single-mode case, but beyond the adiabatic regime, there exists interesting physics and counterintuitive effects[25]. In particular, the intracavity power in a dynamically modulated ring resonator can be completely suppressed even when the input laser is on resonance with the ring, as opposed to the static ring where the intracavity power is maximized on resonance. Thus, a dynamical isolation of the cavity from the input laser can be achieved by operating in the nonadiabatic regime[25,26].

In this paper we present an experimental exploration of such a dynamically modulated ring resonator in the nonadiabatic regime. Using a geometry consisting of a modulated fiber ring cavity coupled with two waveguides, we probe the nonadiabatic regime to demonstrate dynamical isolation and the suppression of drop-port transmission on resonance, based on the aforementioned analytical work[25]. Our experiments can have applications in using dynamic modulation for switching, optical signal processing, waveform synthesis, and frequency conversion of light.

## THEORY

We start by briefly summarizing the relevant theoretical results in Ref. 25. For a ring cavity that is side coupled to two waveguides as shown in Fig. 1(a), the coupled-mode equations for the mode amplitude $\alpha$ are,

$$\frac{d\alpha}{dt} = [i(\omega_0 + \omega(t)) - \gamma]\alpha + \sqrt{\gamma}\, s_{1+}$$

$$s_{1-} = -s_{1+} + \sqrt{\gamma}\alpha(t); \quad s_{2-} = \sqrt{\gamma}\alpha(t),$$

where $s_{j+}$ and $s_{j-}$ represent the input and output amplitudes in the $j$-th port, ($j = 1,2$); $|\alpha|^2$ depicts the electromagnetic energy inside the cavity as carried by a circulating mode; $|s_{j\pm}|^2$ are the respective powers; and $\gamma$ is the decay rate of the cavity into either of the waveguides. $\omega_0$ is the static resonance frequency of the cavity, and $\omega(t)$ is the time-dependent frequency modulation imparted by the electro-optic drive. The countercirculating mode of the resonator has been ignored in the above equations.

For a monochromatic input field $s_{1+} = p_0 \exp[i(\omega_0 + \Delta\omega)t]$ and a cosinusoidal modulation of the resonant frequency $\omega(t) = A_0 \cos \Omega t$, it was shown that the steady state solution for the drop-port output field $s_{2-}$ is [25]

$$s_{2-}(t) = p_0 e^{i(\omega_0+\Delta\omega)t} \sum_n s_n(\Delta\omega) e^{in\Omega t}$$

where,

$$s_n(\Delta\omega) = -\sum_k J_{n+k}\left(\frac{A_0}{\Omega}\right) J_k\left(\frac{A_0}{\Omega}\right) \frac{\gamma}{ik\Omega - i\Delta\omega - \gamma} \quad (1)$$

The transmission at the drop port can be calculated from $T(t) = |s_{2-}|^2/p_0 = \sum_n T_n e^{i\Omega t}$, with $T_n = \sum_m s_m^* s_{n+m}$. Eq. (1), which was derived in Ref. 25, is in agreement with the results in Ref. 27, but its novel consequences in the high-frequency nonadiabatic regime were first discussed in Ref. 25. Eq. (1) was also recently derived in Ref. 28.

In the adiabatic limit predominantly studied in cavity-based modulators,[4] the transmission can be written as,

$$T(t) = T(\omega(t)) = \left|\frac{\gamma}{i(\omega(t) - \Delta\omega) - \gamma}\right|^2 \quad (2)$$

This solution is valid for

$$\frac{A_0}{\Omega} \ll \left(\frac{\gamma}{\Omega}\right)^2 \quad (3)$$

which we call the adiabaticity condition. Since $A_0/\Omega$ was typically of order unity or less in nearly all previous studies of modulated cavities in the single-mode regime, it suffices to say that the adiabaticity is satisfied when the modulation frequency is much smaller than the linewidth of the cavity, or equivalently, the modulation period is much larger than the photon lifetime in the cavity. In this regime, the transmission shows a time variation on a timescale of the modulation frequency.

In the opposite regime, especially when $\gamma \ll \Omega$, the transmission becomes approximately independent of time. In particular, significant transmission only occurs when $\Delta\omega \approx k\Omega$, with $k$ being an integer, in which cases the transmission has the form:

$$T(\Delta\omega \approx k\Omega) = J_k^2\left(\frac{A_0}{\Omega}\right) \left|\frac{\gamma}{i(\Delta\omega - k\Omega) - \gamma}\right|^2 \quad (4)$$

Thus, the transmission can be completely suppressed for zero laser-cavity detuning, i.e., $\Delta\omega = 0$, if the peak phase excursion of the modulation $A_0/\Omega$ is a zero of the Bessel function $J_0$[25]. This is essentially a classical dynamical isolation effect[26]. Note that this is strikingly different from both the static cavity, where the intracavity power and the drop port transmission are maximum for zero detuning, and from the adiabatic regime, where the transmission is determined by the instantaneous frequency of the modulated cavity. In this high-frequency nonadiabatic regime, the transmission can be suppressed even though there are times when the instantaneous frequency of the cavity is on resonance with the incident wave. More generally, this suppression can be realized for a non-zero detuning that is an integer multiple of the modulation frequency ($\Delta\omega = k\Omega$) by choosing the modulation strength to be a zero of the $k$-th order Bessel function $J_k$.

# EXPERIMENTAL SETUP

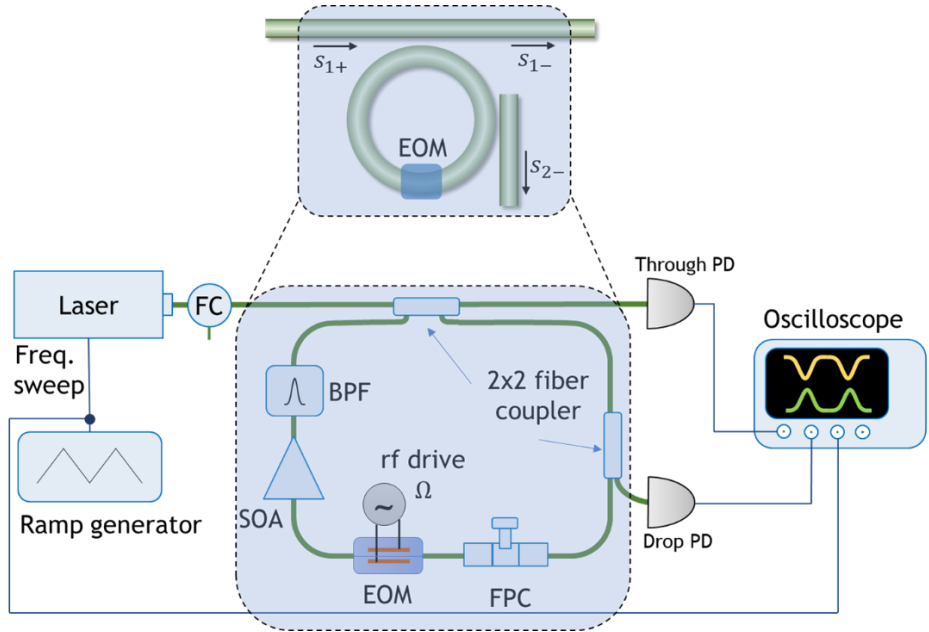

Fig. 1: Experimental setup. The ramp generator is used to sweep the laser wavelength through its frequency sweep input to observe resonances at the through and drop ports of the fiber ring resonator. The rf drive's amplitude and frequency are chosen appropriately to probe either the nonadiabatic or the adiabatic regime. EOM: electro-optic modulator. FPC: Fiber polarization controller. PD: photodiode. BPF: bandpass filter, 26 GHz bandwidth centered at 1542.14 nm. SOA: semiconductor optical amplifier. FC: Fiber circulator.

We experimentally demonstrate the effect of dynamic isolation in an nonadiabatically modulated cavity[25] by incorporating an EOM into a fiber ring resonator, as shown in Fig. 1. As discussed above, to reach the non-adiabatic regime the key is to modulate the cavity at frequencies $\Omega$ much higher than the cavity linewidth $\gamma$. And moreover, in order to be able to demonstrate the effect of dynamic isolation the strength of the modulation $A_0$ needs to be comparable to the modulation frequency $\Omega$. Thus, the experimental setup needs to simultaneously satisfy the following condition: (i) low roundtrip losses, which reduces $\gamma$, (ii) a large modulation frequency $\Omega$, and (iii) a low half-wave voltage, $V_\pi$ for the modulator, to ensure a large modulation amplitude $A_0$ can be achieved for a reasonable voltage. We ensure low roundtrip losses by introducing an optical amplifier to compensate for the insertion loss of intracavity components and connectors[29]. The cavity contains a fiber-coupled Lithium Niobate (LN) phase modulator with a moderately low $V_\pi$ and a sufficiently large maximum modulation frequency of 2 GHz to enable us to reach the nonadiabatic regime and to achieve a large modulation amplitude $A_0$. Note that such all-fiber configurations have been previously explored for realizing synthetic photonic lattices,[30] for generating frequency comb,[10–14] for demonstrating optical Ising machines[31] and for producing nonclassical squeezed light[32].

The detailed experimental setup is shown in Fig. 1. A fiber ring resonator is built from commercially available off-the-shelf components. The ring is coupled to an input fiber through a 2×2, 99:1 fused fiber splitter. We excite the ring with a RIO Orion laser (grade 3, linewidth 2.8 kHz, center wavelength 1542.057 nm). The laser has a frequency sweep input to sweep its wavelength over 3.2 pm (frequency over 400 MHz). A circulator was used before the fiber splitter to monitor any back reflections from the ring cavity. The ring contains a 2 GHz fiber-pigtailed lithium niobate electro-optic phase modulator (JDS Uniphase CATV dual output, x-cut) with an insertion loss of 4 dB and a $V_\pi$ of ~8.5 V. The modulator is driven by a sinusoidal radiofrequency (RF) signal source. By ensuring that the modulation frequency $\Omega/2\pi$ is much less than the FSR of the ring resonator, the effectively single-mode description of the system is valid. To overcome the insertion loss of the modulator and other components, we use a semiconductor optical amplifier (SOA). The SOA can operate with a maximum gain of 12 dB at 1542 nm; however, we operate it with a lower gain to compensate for intracavity losses without inducing lasing. With the intracavity loss compensated, the ring cavity has a finesse of up to 90 due to the input and output coupling. Lower finesse can be obtained by using a lower gain from the amplifier. The amplified spontaneous emission noise from the SOA was filtered using a dense-wavelength-division-multiplexing (DWDM) filter with a 3-dB bandwidth of 26 GHz and a center wavelength of 1542.14 nm. The bandpass filter also serves to inhibit spurious lasing at wavelengths different from the input-coupled light. Note that the bandwidth of this filter is much larger than the FSR of the ring resonator (4.1 MHz or 15 MHz), so that its response is flat over the frequency range of interest here. A fiber polarization controller was used to align the input polarization to the principal axis of LiNbO$_3$ in the EOM. A drop port is incorporated using a second 99:1 fiber splitter. We exclusively use angle-polished fiber connectors (APC) throughout the setup to minimize back reflections and prevent exciting the countercirculating mode in the ring. The through port and drop port powers are monitored with InGaAs photodiodes with electrical bandwidths greater than 5 GHz. We also use a proportional-integral (PI) feedback loop (not shown in Fig. 1) to lock the laser wavelength to a resonance of the ring cavity by actuating on the laser's frequency modulation input. Additionally, a fraction of the input laser light is frequency shifted using an acousto-optic modulator and then heterodyned with the cavity output if needed, to perform spectrally resolved field measurements ($s_n$) instead of power measurements ($|s_{1-}(t)|^2, and\ |s_{2-}(t)|^2$ ).

The adiabatic and nonadiabatic regimes can both be probed using this setup by appropriately choosing a combination of the modulation amplitude, the modulation frequency and the optical amplifier gain to control $A_0$, $\Omega$ and $\gamma$ respectively. For the adiabatic regime, we use a cavity with a linewidth of 200 kHz and a modulation frequency $\Omega/2\pi = 10$ kHz, which allows for a large modulation amplitude of $A_0/\Omega \leq 10$ while still satisfying Eq. (3). The nonadiabatic regime is explored using a cavity with a similar linewidth of 270 kHz, but with a much larger modulation frequency of 1.3 MHz, enabling us to violate the adiabaticity condition in Eq. (3) with very small modulation amplitudes $A_0/\Omega \geq 0.01$.

## RESULTS

**Adiabatic regime**

We first study the cavity in the adiabatic regime that is typically associated with cavity EOMs and observe good agreement between the analytical predictions and experimental results, as depicted in Fig. 2. For this regime, the cavity was implemented to have a linewidth $2\gamma = 2\pi \cdot 200$ kHz and an FSR of 4.1 MHz (Finesse $\approx 20$). The low finesse is beneficial for probing the adiabatic regime and was achieved by reducing the gain of the optical amplifier.

For a static cavity, the maximum transmission occurs for zero detuning, and falls off with increasing values of detuning, as depicted by the insets in Fig. 2. For the dynamically modulated cavity, we operate in the adiabatic regime by applying a modulation signal such that the peak phase excursion of the modulation is $A_0/\Omega = 4.3$, and $\Omega/2\pi = 10$ kHz. The adiabaticity condition (Eq. (3)) is well satisfied by these parameters: $A_0/\Omega = 4.3 \ll (\gamma/\Omega)^2 = 100$. Here the slope of the static Lorentzian for a certain detuning determines the peak-to-peak swing of the transmission, as shown by the insets in Fig. 2 and the corresponding transmission signals for various detunings (Fig 2(a) – (d)). In this regime, the transmission varies as a function of time, and such a variation can be well understood using Eq. (2), where the transmitted power at any instant is determined by the instantaneous resonance frequency. At zero detuning (Fig. 2a), the transmission varies at twice the frequency of the modulation due to the vanishing slope and the quadratic response near resonance. With larger detuning (Fig. 2c and d), the transmission varies at the frequencies of the modulation. The magnitude of the transmission variation decreases as the detuning increases. These behaviors are expected for a standard electro-optic cavity modulator.

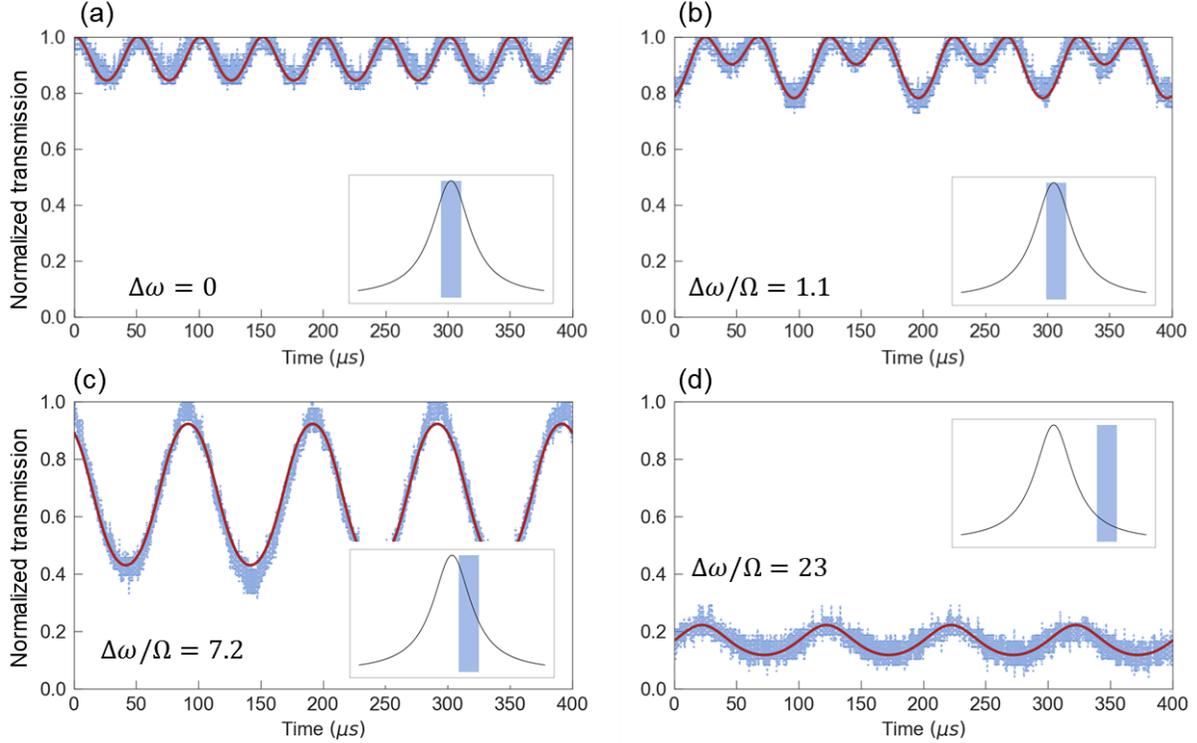

Fig. 2: Normalized transmission through the cavity in the adiabatic regime for various values of the input laser detuning $\Delta\omega/\Omega = 0$ (a), 1.1 (b), 7.2 (c), and 23 (d). The red solid lines are analytical curves based on Eq. (2) for the corresponding experimental data shown by blue dots. The insets in each figure show the static Lorentzian response of the cavity and the range over which the modulation sweeps the instantaneous resonance frequency. Here, the modulation frequency $\Omega = 2\pi \cdot 10$ kHz, much smaller than the linewidth $2\gamma = 2\pi \cdot 200$ kHz. Phase modulation amplitude $A_0/\Omega = 4.3$.

**Nonadiabatic, high-frequency regime**

In contrast to the adiabatic regime presented above, we now experimentally observe the suppression of drop-port transmission and consequent dynamical input isolation in the nonadiabatic regime, as shown in Fig. 3. For probing this regime, a shorter cavity was used ($2\gamma = 2\pi \cdot 270$ kHz, FSR = 15 MHz). A modulation frequency of $\Omega/2\pi = 1.3$ MHz ensured that the system operated in the high-frequency regime ($\Omega \gg \gamma$). In this regime, nonadiabaticity is achieved even for a low modulation amplitude $A_0/\Omega > (\gamma/\Omega)^2 = 0.01$. To observe the transmission spectrum, we sweep the input laser frequency across a ring resonance slowly using a 200 Hz triangular ramp applied to fine tune the input frequency, and record the total transmitted power to measure the steady state response of the modulated cavity at each input frequency. For $A_0/\Omega = 0$, the transmitted power exhibits a Lorentzian dependency with respect to the input frequency. On the other hand, when $A_0/\Omega$ is large, the transmitted power exhibits a peak whenever the input detuning is near $\Delta\omega = k\Omega$, as shown by the red curve in Fig. 3, in agreement with Eq. (4). Moreover, when $A_0/\Omega \approx 2.4$, which approaches a zero of the Bessel function, as shown in Fig. 3, the transmission nearly completely vanishes for an on-resonance input with $\Delta\omega = 0$. Such an absence of transmission signifies the effect of dynamic isolation: in the presence of modulation the cavity is no longer excited by an on-resonance input and hence is isolated from the input. As $A_0/\Omega$ varies from 0 to 2.4, the power transmission coefficient for an on-resonance input varies from 1 to 0. In this system, the amplitude of the sinusoidal modulation of the resonant frequency thus provides a switching mechanism based on the dynamical isolation effect.

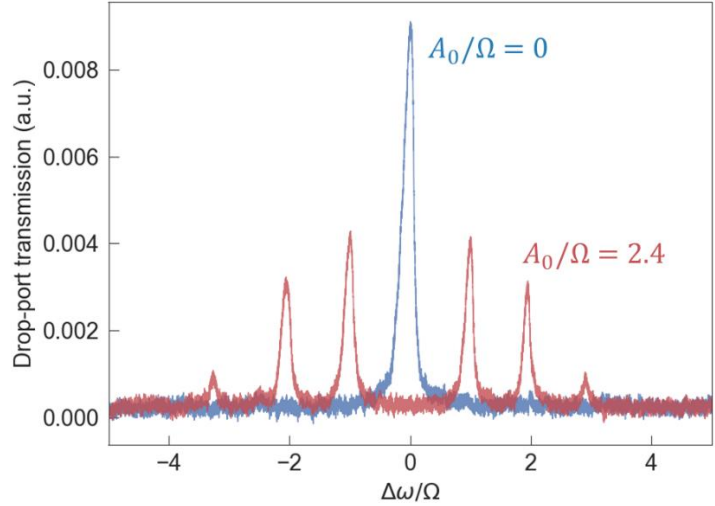

Figure 3: Nonadiabatic, high-frequency regime. Variation of the drop-port transmission with detuning for no modulation, $A_0/\Omega = 0$ (0 $V_{pp}$) and for $A_0/\Omega = 2.4$ ($\equiv 8.6\ V_{pp}$), which is close to the zero of the Bessel function $J_0$. The transmission at zero detuning is suppressed by applying this modulation, resulting in dynamical input isolation. Cavity linewidth $2\gamma = 2\pi \cdot 270$ kHz, $\Omega = 2\pi \cdot 1.3$ MHz, FSR = 15 MHz. These parameter values satisfy both the constraints of single-mode operation ($\Omega/2\pi \ll$ FSR), and nonadiabatic high-frequency modulation ($\Omega \gg \gamma, A_0/\Omega \geq (\gamma/\Omega)^2 = 0.01$).

More generally, we can suppress the transmission at a non-zero detuning that is an integer multiple of the modulation frequency, by choosing $A_0/\Omega$ to be a zero of the Bessel function of the appropriate order. One could also vary the modulation frequency to achieve a suppression of transmission at a desired detuning from the cavity in this manner. We illustrate this in Fig. 4(a) by mapping out the drop-port transmission with detuning $\Delta\omega$ for increasing values of the modulation amplitude $V_{pp}$. $A_0$ in our experiments is determined by the peak-to-peak amplitude of the voltage applied to the EOM. Hence, to map out a larger range of $A_0/\Omega$ within the maximum achievable $V_{pp}$ of 10 V, we use a lower modulation frequency $\Omega/2\pi = 0.8$ MHz. The drift in the spectrum between different modulation amplitudes was factored out. We see that with an increasing modulation amplitude, the transmission can be suppressed at $\Delta\omega/\Omega = 0, \pm 1$ for $V_{pp} \approx 5.8, 9.3$ respectively, corresponding respectively to $A_0/\Omega = 2.4, 3.8$, which are close to the zeros of the Bessel function. The zero for $\Delta\omega/\Omega = \pm 2$ and higher orders is not attained due to the maximum achievable RF amplitude delivered by signal generator. Fig 4(b)-(d) shows line cuts from Fig. 4(a) for fixed detuning values corresponding to $\Delta\omega/\Omega = 0$ (blue), 1 (red) and 2 (green). The solid lines are fits based on the analytical prediction of Eq. (2) and show good agreement with the experimental data.

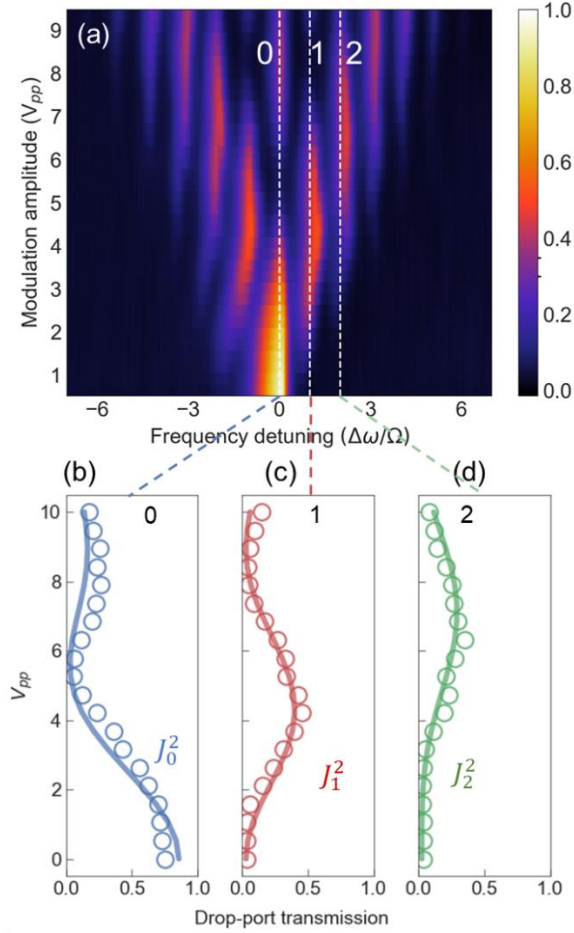

Figure 4: (a) Variation of the drop port-transmission with detuning and modulation amplitude. The drop-port transmission has been normalized to its maximum in the absence of modulation. (b)-(d) Line-outs along the dashed vertical lines in (a). The circles represent experimental data. The solid lines are fits to Eq. (1) for $\Delta\omega/\Omega$ = 0, 1 and 2, which reduce to squares of the Bessel functions of the first kind and of the corresponding order. Zeroes of $J_0$ and $J_1$ are seen in the data. The drift in the spectrum on successive scans of increasing modulation amplitude was subtracted out.

Lastly, we show a direct measurement of the output field by heterodyning the drop port signal with a frequency shifted version of the input laser, to recover the spectral components $s_n(\Delta\omega)$ in Eq. (1). For this experiment, about 90% of the input laser is split off and sent through a free-space acousto-optic modulator (AOM). The AOM is driven by an RF signal at 500 MHz, and the first order diffraction from the AOM produces a beam that is up-shifted by 500 MHz in frequency. The output of the AOM is coupled back into fiber and mixed with the drop port signal at a 50:50 fiber beam splitter. The spectral components of this signal around 500 MHz yield the output field components $s_n(\Delta\omega)$. This process is repeated while scanning the input laser frequency, resulting in the plot in Fig. 5(a). The experimental data agrees very well with its analytically calculated counterpart based on Eq. (1) and outlined in Ref. 25. For example, $|s_{-1}|^2$ has peaks predominantly at $\Delta\omega/\Omega = 0, 1$, whereas $|s_2|^2$ has peaks at $\Delta\omega/\Omega = 0, -1, -2$.

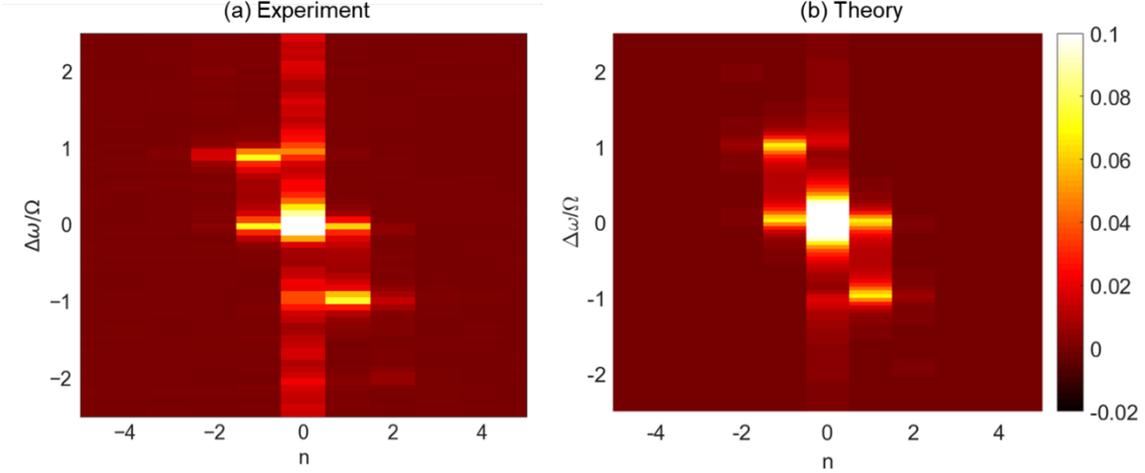

Figure 5: Spectral components $|s_n(\Delta\omega)|^2$ of the output field, measured experimentally using a heterodyne technique (a), and compared with the analytically calculated values (b) using Eq. (1). $A_0/\Omega = 0.6$, $\Omega/2\pi = 1$ MHz.

## CONCLUSION AND FINAL REMARKS

We have experimentally demonstrated the theoretical predictions in the analytical work of Minkov et al.[25], using a fiber ring resonator with an electro-optic phase modulator inside the cavity. We observed dynamical input isolation and suppression of drop-port transmission when the cavity was on resonance with the input laser, for appropriate parameters of the modulation. We also verified the behavior of this cavity in the adiabatic regime.

Similar demonstrations can be realized with on-chip photonic technologies, either by compensating the high insertion loss of chip-based EOMs with a gain medium inside the cavity, or perhaps more practically by using low-loss EOMs such as those recently reported in lithium niobate[33–39]. For realizing gain on chip, several approaches exist, such as incorporating III-V-based amplifiers,[40–45] which requires electrical pumping, or using erbium doping or parametric gain, which require optical pumping[46–53]. The approach of utilizing on-chip gain might be especially important for implementing nonadiabatically modulated cavities in silicon. This is because silicon electro-optic modulators are based on the plasma dispersion effect, where an increasing modulation strength is inevitably concomitant with increasing loss[54].

One could envision extending the current platform for frequency conversion and signal optimization by combining multiple harmonic modulations in tandem. Several theoretical proposals extend the dynamic modulation approach to an array of coupled cavities for realizing photonic gauge potentials, synthetic dimensions, topological photonics and for controlling light transport,[15–18,20,55–60] and our experimental platform is ripe for realizing these proposals in an all-fiber configuration. In a broader context of dynamical decoupling, the concept of modulating an open system at a rate faster than the timescale involved in the system-reservoir interaction has been proposed as a means of isolating it from the environment and suppressing decoherence[61,62]. This can be implemented and studied in a setup similar to the nonadiabatically modulated cavity we have presented, both for classical and for quantum optics applications.

## AUTHOR INFORMATION

**Corresponding author**

*Email: shanhui@stanford.edu

**Notes**

The authors declare no competing financial interest.

## ACKNOWLEDGMENTS

This work is supported by a Vannevar Bush Faculty Fellowship (Grant No. N00014-17-1-3030) from the U. S. Department of Defense, and by a MURI grant from the U. S. Air Force Office of Scientific Research (Grant No. FA9550-17-1-0002).